\documentclass[twocolumn]{aa} 
\pdfoutput=1

\usepackage{graphicx}
\usepackage{txfonts}
\usepackage{natbib}
\bibpunct{(}{)}{;}{a}{}{,} 

\newcommand{\rSun}{r/R_{\odot}}
\newcommand{\RSun}{R_{\odot}}
\newcommand{\Done}  {${\sf D}_1$}
\newcommand{\Dtwo}  {${\sf D}_2$}
\newcommand{\Dthree}{${\sf D}_3$}
\newcommand{\Dfour} {${\sf D}_4$}
\newcommand{\Dfive} {${\sf D}_5$}

\begin{document}
%
\title{Sensitivity of helioseismic gravity modes to the dynamics of the solar core}

   \author{S. Mathur\inst{1} \and
                A. Eff-Darwich\inst{2,3} \and 
                R.A. Garc\'\i a\inst{1} \and
   		S. Turck-Chi\`eze\inst{1}
	}
   \offprints{smathur@cea.fr}
   \institute {Laboratoire AIM, CEA/DSM -- CNRS - Universit\'e Paris Diderot -- IRFU/SAp, 91191 Gif-sur-Yvette Cedex, France
   \and Departamento de Edafolog\'\i a y Geolog\'\i a, Universidad de La Laguna, Tenerife, Spain
   \and Instituto de Astrof\'\i sica de Canarias, 38205, La Laguna, Tenerife, Spain
        }

   \date{Received 2007; accepted 2008}

 \abstract
    {The dynamics of the solar core cannot be properly constrained through the analysis of acoustic oscillation modes. Gravity modes are necessary to understand the structure and dynamics of the deepest layers of the Sun. Through recent progresses on the observation of these modes -- both individually and collectively -- new information could be available to contribute to inferring the rotation profile down inside the nuclear burning core.}
   {To see the sensitivity of gravity modes to the rotation of the solar core. We analyze the influence of adding the splitting of one and several g modes to the data sets used in helioseismic numerical inversions. We look for constraints on the uncertainties required in the observations in order to improve the derived core rotation profile.}
   {We compute forward problems obtaining three artificial sets of splittings derived for three rotation profiles: a rigid profile taken as a reference, a step-like and a smoother profiles with higher rates in the core. We compute inversions based on Regularized Least-Squares methodology (RLS) for both artificial data with real error bars and real data. Several sets of data are used: first we invert only p modes, then we add one and several g modes to which different values of observational uncertainties (75 and 7.5 nHz) are attributed. For the real data, we include g-mode candidate, $\ell$=2, n=-3 with several splittings and associated uncertainties.}
   {We show that the introduction of one g mode in artificial data improves the rate in the solar core and give an idea on the tendency of the rotation profile. The addition of more g modes gives more accuracy to the inversions and stabilize them. The inversion of real data with the g-mode candidate gives a rotation profile that remains unchanged down to 0.2 $\RSun$, whatever value of splitting we attribute to the g mode.}
  {}


   \maketitle
   
\section{Introduction}

The Sun is a magnetic star and it is now well recognized that the dynamical processes occurring in the 
 solar interior are linked to the activity of the visible external layers (For a review of solar and stellar activity see \citet{2000ssma.book.....S}).  The fact that the Sun is still
 active today, even at the present low rotation
 rate (in comparison to  young stars), implies that the initial magnetic fields are maintained, regenerated or 
amplified through  dynamo effects which are induced by fluid motions within the star, namely 
rotation, convection and/or meridional circulation (support for this comes from observations, e.g., \citet{1987ARA&A..25..271H}). In this sense, it is necessary to reconstruct the solar 
internal rotational profile from the surface down to the core, to properly understand the magnetic activity 
of the Sun. Over the past decade, increasingly accurate helioseismic observations from
ground-based and space-based instruments have given us a reasonably good
description of the dynamics of the solar interior 
\citep[][ and references therein]{1998ApJ...505..390S,2000ApJ...541..442A,ThoJCD2003}. Helioseismic inferences have
 confirmed that the differential
rotation observed at the surface persists throughout the convection
zone. There appears to be very little, if any, variation of the rotation rate
with latitude in the outer radiative zone ($0.4 > \rSun > 0.7$). The rotation rate is almost constant ($\approx 430$ nHz) in this region which is separated from the region of differential rotation by a narrow shear layer ---known as the tachocline \citep{1992A&A...265..106S,1998A&A...330.1149C} ---.

The rotation profile of the Sun is also connected to different aspects of the structure and dynamics of the star.
 This is the reason why the rotation rate is needed to estimate the circulation
 and shear instabilities which are responsible for the redistribution of chemical elements \citep{ThoJCD2003}. Moreover, 
the redistribution of angular momentum through the coupling between the turbulent convection and the
 rotation contributes to the strong differential rotation in the convective zone \cite[e.g.][]{1998ApJ...505..390S} and hence, to the dynamo effect 
that is thought to be responsible for the 11-year activity cycle 
\citep[][ and references therein]{2004ApJ...614.1073B,2007ApJS..170..203G} and the evolution of the Hale solar cycle
  \citep{2006ApJ...649..498D}. 

Although the helioseismic inferences in the radiative zone are not as precise as those found in the convective region, it could be confirmed that the rotation rate is flat and rigid down to approximately 0.4 $\RSun$. At least three processes have been proposed to explain this flatness of the rotation profile in the radiative zone but without great success: the redistribution of angular momentum by the effect of differential rotation which does not succeed in producing a completely flat profile \citep{1997A&A...317..749T}; the effect of some magnetic fossil field instabilities that could flat the profile but they have not yet been found \citep[][]{2002A&A...381..923S,2005A&A...440L...9E} and finally, some transport of angular momentum by internal gravity waves \citep[][]{2002ApJ...574L.175T,2005Sci...309.2189C,Turck-ChiezeTalon2007}. 
 A general formalism has been developed recently to take into account all these different
 processes \citep[][]{2004A&A...425..229M,2005A&A...440..653M}, but more accurate observations of the solar rotation profile
 are needed to constrain  the theoretical picture.

 \begin{figure}[htb*]
\begin{center}
\includegraphics[height=7cm]{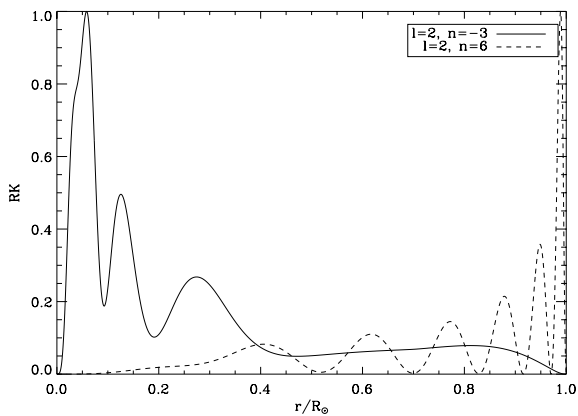}
\caption{Rotational kernels for a p mode ($\ell$ = 2, n=6) and the g-mode candidate ($\ell$ = 2, n=-3).}
\label{rot_k}
\end{center}
\end{figure}

The analysis of the rotation profile at deeper layers (below 0.4 $R_{\odot}$) and therefore, inside the solar burning core
(where  more than half of the solar mass is concentrated) could only be carried out with 
 a few tens of p modes -- the low-degree modes ($\ell \le$3) --. Indeed, since the dawn of helioseismology when the works by 
\citet{1981Natur.293..443C,1982Natur.299..704C} led to the conclusion that the solar core rotates from 2 to 9
 times faster than the surface rate, several groups have published different estimations of the rotation rate in the 
solar core -- using acoustic modes -- with contradictory results 
\citep[][]{1994ApJ...435..874J,ElsHow1995,1995ASPC...76...24F,1996SoPh..166....1L,ChaEls2001}. The importance of the low order p modes (below 2.3 mHz) to properly establish the profile below 0.4 $\RSun$ has been shown \citep{CouGar2003}. Gravity modes having large sensitivities to
 the solar core, will significantly contribute to establish the actual dynamical conditions of the core. This is illustrated in Fig.~\ref{rot_k} representing the rotational kernel for a g mode ($\ell$ = 2, $n$ = -3) and for a p mode ($\ell$ = 2, $n$ = 6). This g mode is mostly sensitive to the region below 0.2 $\RSun$ whereas the p mode is sensitive to the region above 0.4 $\RSun$. It shows the importance of g modes compared to p modes for having access to the rotation of the core.

 The advent of the
 new millennium saw the burgeoning of the g-mode research based on the quality and accumulation of 
helioseismic data. 
In 2000, \citeauthor{AppFro2000} looked for individual spikes above 150 $\mu$Hz  in the power spectrum with more than
 90\% confidence level that the signal was not pure noise.  Although they could not identify any g-mode signature, an 
upper limit of
 their amplitudes could be established: at 200 $\mu$Hz, they would fall below 10 $mm s^{-1}$ in velocity, and below 0.5 
parts per million in intensity. Later, in 2002, \citeauthor{GabBau2002}, using the same  statistical approach,
found a peak that could be interpreted  as one component of the $\ell$=1, $n$=1 mixed mode. 

A different approach based on the search of multiplets and recurrent signals in 
time \citep{GarTurck1997,PalleGarcia1997,1998ESASP.418..549T} have been applied to GOLF\footnote{Global Oscillations at Low Frequency 
\citep{GabGre1995}}/SoHO\footnote{Solar and Heliospheric Observatory \citep{DomFle1995}} velocity time series 
\citep{GarSTC2005}. Some time-coherent patterns 
were found in the signal \citep{GabSTC1999}, thus they could be potentially considered as g modes.  
\cite{STCGar2004} applied this technique to high-frequency multiplets in hope of reducing 
the detection threshold while maintaining the same confidence level. These authors found several patterns attributed to g-mode signals and, in particular, one was considered as a candidate for the mode $\ell=2, n=-3$. In fact, \citet{CoxGuz2004} postulated theoretically that this mode could be 
the one with the largest amplitude at the solar surface. This candidate is still present in the analysis of longer 
time series \citep[][ and references therein]{2007ApJ...668..594M}. Finally, the measurement of a signal that could be attributed to the separation in period of the dipole gravity modes and the comparison with solar models fosters a faster rotation rate in the core than the rest of the radiative zone \citep{2007Sci...316.1591G}.

In this work, we will study how the inferences about the solar core rotation profile  could be improved by including 
gravity modes. We will study the effect of adding either one (the candidate $\ell$=2 n=-3) or several g modes 
in the data set that will be inverted to infer the rotational profile. The effect of the observational uncertainties
on the derived rotational rate will also be analyzed, as well as the introduction of the g-mode candidate in real 
p-mode data sets.






\section{Methodology.}

Helioseismic inferences on the internal rotation rate of the Sun are carried out through 
numerical inversions of the functional form of the perturbation in frequency, $\Delta \nu _{n \ell
m}$, induced by the rotation of the Sun, $\Omega(r,\theta)$ and given by
\citep[see derivation in][]{bi46}:
\begin{equation} 
\Delta \nu _{n \ell m} = \frac{1}{2\pi}\int_0^R \int_0^{\pi}
K_{n \ell m}(r,\theta)\Omega(r,\theta)drd\theta + \epsilon_{n \ell m}
\label{eq:equation1} 
\end{equation}

  The perturbation in frequency, $\Delta \nu _{n \ell m}$ with error
$\epsilon_{n \ell m}$, that corresponds to the rotational component of the
frequency splittings, is given by the integral of the product of a sensitivity
function, or kernel, $K_{n \ell m}(r,\theta)$ with the rotation rate,
$\Omega(r,\theta)$, over the radius, $r$, and the co-latitude, $\theta$. The
kernels, $K_{n \ell m}(r,\theta)$, are known functions of solar models. 

Equation \ref{eq:equation1} defines the forward problem for the solar interior 
rotation rate through global helioseismology, since it is possible to calculate 
estimates of the frequency splittings, $\Delta \nu _{n \ell m}$, that correspond 
to a given solar rotation rate,  $\Omega(r,\theta)$.

\begin{figure}[h]
\begin{center}
\includegraphics[width=9cm]{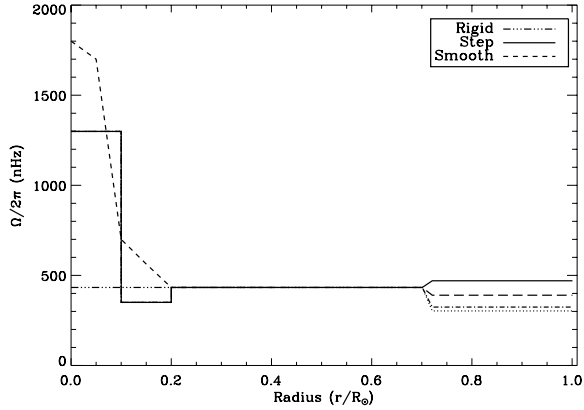}
\caption{Artificial rotation profiles for the solar interior as explained in the text and used in the computation of the artificial data sets. The three artificial profiles have the same behavior in the convective zone. They incorporate latitudinal variations in the convection zone to mimic the actual rotation profile of the Sun. The plotted latitudes are 0$^o$ (solid line), 30$^o$ (dashed line), 60$^o$ (dotted line)  and 75$^o$ (dashed-dotted line). }
\label{profil}
\end{center}
\end{figure}

 The latter equation also defines a classical inverse problem for the
sun's rotation. The inversion of this set of $M$ integral equations -- one for
each measured $\Delta \nu _{n \ell m}$ -- allows us to infer the rotation rate
profile as a function of radius and latitude from a set of observed rotational
frequency splittings (hereafter referred to as splittings). The inversion method we 
have used is based on the regularized least-squares
methodology (RLS). The RLS method requires the discretization of the integral
relation to be inverted. In our case, Eq.~1 is transformed
into a matrix relation
\begin{equation}
  D = A x + \epsilon \label{eq:equation2}
\end{equation}
where $D$ is the data vector, with elements $\Delta \nu _{n \ell m}$ and
dimension M, $x$ is the solution vector to be determined at $N$ tabular
points, $A$ is the matrix with the kernels, of dimension $M \times N$ and
$\epsilon$ is the vector containing the errors in $D$. 

The RLS solution is the one 
that minimizes the quadratic difference
$\chi^2=|Ax-D|^2$, with a constraint given by a smoothing matrix, $H$,
introduced in order to lift the singular nature of the problem \citep[see, for
instance,][]{Eff-Darwich1997}. Hence, the function $x$ is approximated
by 

\begin{equation}
  x_{\rm est} = (A^TA + \Lambda H)^{-1}A^TD  \label{eq:equation3}
\end{equation}

\noindent where $\Lambda$ is a vector defining how much regularization is applied to each 
 point $x_{i}$ of the inversion mesh, as introduced in \cite{Eff-Darwich1997}.

As a by-product of the inversion methodology, we could replace $D$ from Eq.~\ref{eq:equation2} 
to obtain

\begin{equation}
  x_{\rm est} = (A^TA + \Lambda H)^{-1}A^TAx \stackrel{\mathrm{def}}{=} Rx  
\label{eq:equation4}
\end{equation}
hence
\begin{equation}
R = (A^TA + \Lambda H)^{-1}A^TA \label{eq:equation5}
\end{equation}

\noindent The matrix $R$, that combines forward and inverse mapping, is referred to as
the resolution or sensitivity matrix, while the $i_{th}$ row is referred as the resolution kernel for
 the estimation of $x_{i}$ \citep{eff2007}. The diagonal
elements $R_{ii}$ state how much of the information is saved in the model
estimate and may be interpreted as the resolvability or sensitivity of
$x_{i}$. In this sense, it could be possible to use $R_{ii}$ to see the effect 
of modifying the mode sets used in the inversion on the resolvability 
of each point of the inversion mesh.


\begin{figure}[htb*]
\begin{center}
\includegraphics[width=9cm]{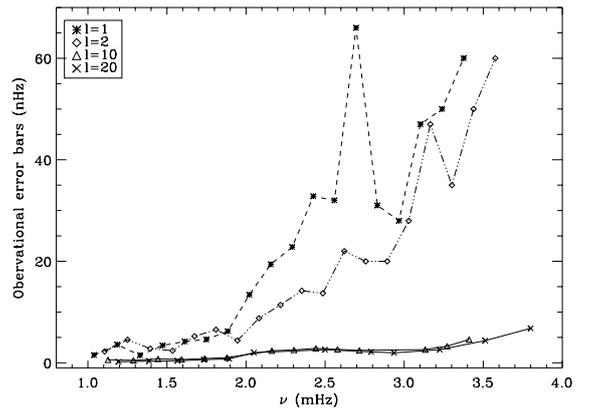}
\caption{Observational error bars of p-mode splittings for degrees $\ell$=1,2, 10 and 20 as a function of the central frequency of the mode. Each degree is represented by a symbol as explained in the legend of the figure.}
\label{real}
\end{center}
\end{figure}

A theoretical analysis was carried out in order to determine the effect of
the addition of g modes on the derivations of the solar rotation rate
of the burning core.  Different artificial data sets have been calculated using
Eq.~\ref{eq:equation1} and three artificial rotation rates $\Omega(r,\theta)$ that are shown in Fig.\ref{profil}. 
 They all have a differential rotation in the convection zone and a rigid rotation from 0.7 down to 0.2 $\RSun$ equal 
to $\Omega_{rz}$=433 nHz. In the first profile -- the rigid profile --, that is our reference profile, the flat and rigid rotation includes the core. The second profile -- the step profile --, is a step-like profile having a rate 3 times larger than the rest of the radiative zone below 
0.1 $\RSun$ and a rate of 350 nHz in the region 0.1-0.2 $\RSun$. Though this profile has unrealistic steep changes, it 
is useful to check the quality of the inversion as these steep profiles are difficult to reproduce. The rotational 
rate for the third profile -- the smooth profile -- increases gradually from 433 nHz at 0.2 $\RSun$ reaching 1800 nHz in the centre, being in this sense compatible with the latest theoretical studies. 

\begin{figure}[ht*]
\begin{center}
\includegraphics[height=7cm]{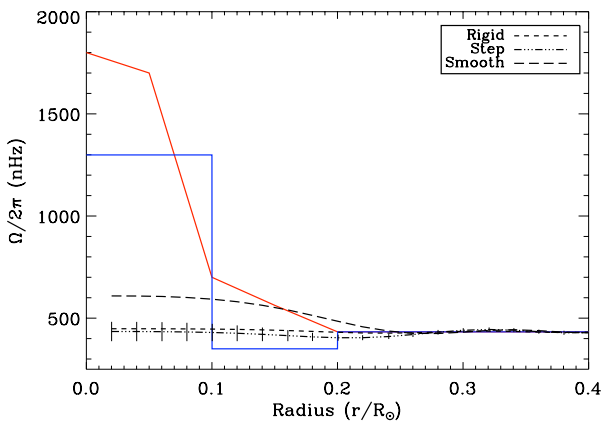}
\caption{Equatorial rotation profiles below 0.4 $R_\odot$ reconstructed with the p modes (Set $D_1$) for the rigid profile (dotted line), the step profile (triple dotted-dashed line) and the smooth profile (dashed line). For the sake of clarity we have plotted the error bars in the step profile only. The continuous blue, red and small dashed lines are respectively the step, smooth and rigid artificial rotation profiles.}
\label{rot_p}
\end{center}
\end{figure}

The different artificial data sets correspond to different mode sets, as explained in
Table \ref{table1}. The observational uncertainties for p modes (see Fig. \ref{real})  were calculated through Principal Component Analysis of the mode sets 
extracted from a sample of 728 days-long MDI\footnote{Michelson Doppler Imager \citep{1995SoPh..162..129S}} 
time-series \citep{kor2005} for p modes with degrees ranging 
from $\ell$=4 to 25, whereas for  $\ell$=1, 3 modes, the uncertainties were extracted from a combined GOLF-MDI time series 
\citep{GarCor2004}. 
The degree range of all data sets spans from $\ell$=1 to 25, however the frequency range 
of the artificial data sets depends on the degree of the mode, ranging from 1 to 2.3 mHz for $\ell$=1 to 3  and from 
1 to 3.9 mHz for $\ell$=4 to 25. As it is illustrated in Fig. \ref{real}, the uncertainties above 2.3 mHz for low degree 
modes are very large,  since it is more difficult to estimate the splittings as a consequence of the blending between the multiplet components of the modes due to the reduction in their life times \citep[][]{BerVar2000,GarReg2001,GarCor2004,ChaEls2002,CouGar2003}. Up to eight different g modes have been 
used in this work, four $\ell$=1 (with frequencies down to 100 $\mu$Hz) and four $\ell$=2 (with frequencies down to 
150 $\mu$Hz) which are the modes with 
the highest predicted amplitudes \citep[][]{KumQua1996,ProBer2000}. Since g modes have not yet been characterized, 
different theoretical uncertainties have been used during the inversion process. Indeed the first value of 75 nHz for the uncertainty on the g-mode splitting, corresponds to the tolerance in the search algorithm used by \citet{STCGar2004} and is related to a possible shift of the multiplet components due to the presence of  a central magnetic field. \citet{2007MNRAS.377..453R} have already shown that this would shift the central frequencies of g modes in this region by such amount. The other value of 7.5 nHz is a typical uncertainty that could be obtained by fitting dipolar modes with $\sim$ 4 years of data and could be a good example of what we could measure in the near future with the next generation of instruments.

\begin{table*}[ht]
  \begin{center} 
\caption{Description of the artificial data sets used to study the sensitivity
of $p$ and $g$ modes to the dynamics of the solar core.}
\vspace{1em}
\renewcommand{\arraystretch}{1.2}
\begin{tabular}[h]{lccccc}
\hline ~        &   \multicolumn{4}{c}{Freq.~range (mHz)} \\
\hline Data set &  $g$ modes  & Uncertainty on $g$ modes (nHz) & $p$ modes $\ell=1, 3$  &  $p$ modes  $\ell > 3$ \\\hline
  Set {\Done} &  -  & -  & $1\le\nu\le2.3  $ & $1\le\nu\le3.9$  \\
 Set {\Dtwo} &  $\ell=2$, $n=-3$ & 75 & $1\le\nu\le2.3$ & $1\le\nu\le3.9$  \\
 Set {\Dthree} &   $\ell=2$, $n=-3$ & 7.5 & $1\le\nu\le2.3$ & $1\le\nu\le3.9$  \\
 Set {\Dfour} & $\ell=1$, $n=-2$ to $-5$ and  $\ell=2$, $n=-3$ to $-6$ & 75 & $1\le\nu\le2.3$ & $1\le\nu\le3.9$  \\
 Set {\Dfive} & $\ell=1$, $n=-2$ to $-5$ and  $\ell=2$, $n=-3$ to $-6$ & 7.5 & $1\le\nu\le2.3$ & $1\le\nu\le3.9$ \\\hline \\
      \end{tabular}
    \label{table1}

  \end{center}
\end{table*}

\section{Results}

\subsection{Inversions of artificial data}

A set of numerical inversions were carried out to study the effect of adding g modes on the derivation of the rotation rate of the solar core. The analysis of the inversion results were complemented with the study of the resolution kernels of the inversions and the direct comparison of the sets of frequency splitting used. 

\begin{figure}[htb*]
\begin{center}
\includegraphics[width=9cm]{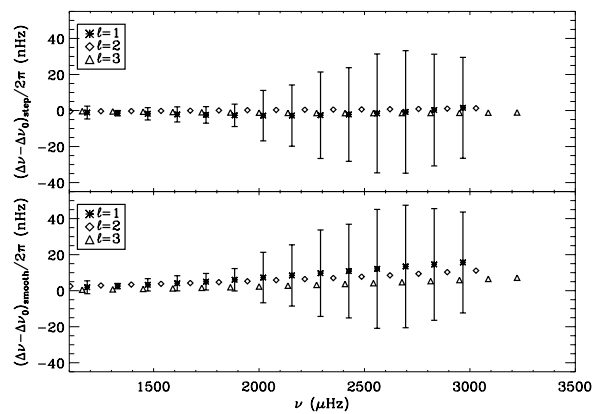}
\caption{Difference of the p-mode splittings between the step profile and on the one hand the rigid profile (top)  and on the other hand, the smooth profile (bottom). }
\label{dif_split1}
\end{center}
\end{figure}

\begin{figure}[ht*]
\begin{center}
\includegraphics[height=7cm]{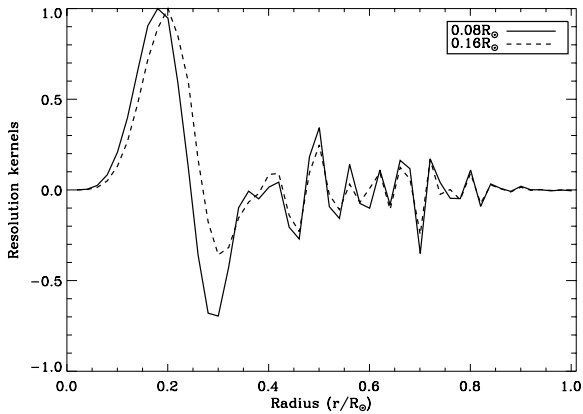}
\caption{Resolution kernels computed in the inversion of the set $D_1$ containing only p modes and calculated at two radii: 0.08 $\RSun$ (solid line) and 0.16 $\RSun$ (dashed line).}
\label{ave_p}
\end{center}
\end{figure}

The inversion of the available p-mode splittings (see Fig.\ref{rot_p}), as those included in set {\Done}, reveals that it is not possible to recover any of the three artificial rotation profiles below 0.2 $\RSun$ (see Fig.\ref{profil}). This result is also illustrated by the comparison of the splittings calculated from the rigid, step and smooth profiles (see Fig.\ref{dif_split1}), since such differences fall below 1 nHz, being the present level of uncertainties for these splittings above this value. The resolution kernels for these inversions (Fig.\ref{ave_p}) also confirm the lack of sensitivity below 0.2 $\RSun$, since it is not possible to properly locate and recover the resolution kernels below 0.2 $\RSun$.

\begin{figure}[ht*]
\begin{center}
\includegraphics[height=7cm]{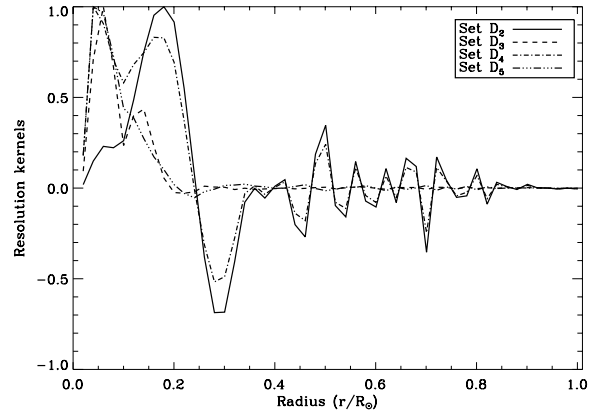}
\caption{Resolution kernels computed in the inversion of the set $D_2$ (i.e. including the g mode $\ell$=2 $n$=-3 with an error bar of 75 nHz), $D_3$(i.e., including the g mode with an error bar of 7.5 nHz), $D_4$ (i.e., including eight g modes with an error bar of 75 nHz) and $D_5$ (i.e., including eight g modes with an error bar of 7.5 nHz) at 0.08 $\RSun$.}
\label{ave_0_08}
\end{center}
\end{figure}

\begin{figure}[htb*]
\begin{center}
\includegraphics[height=7cm]{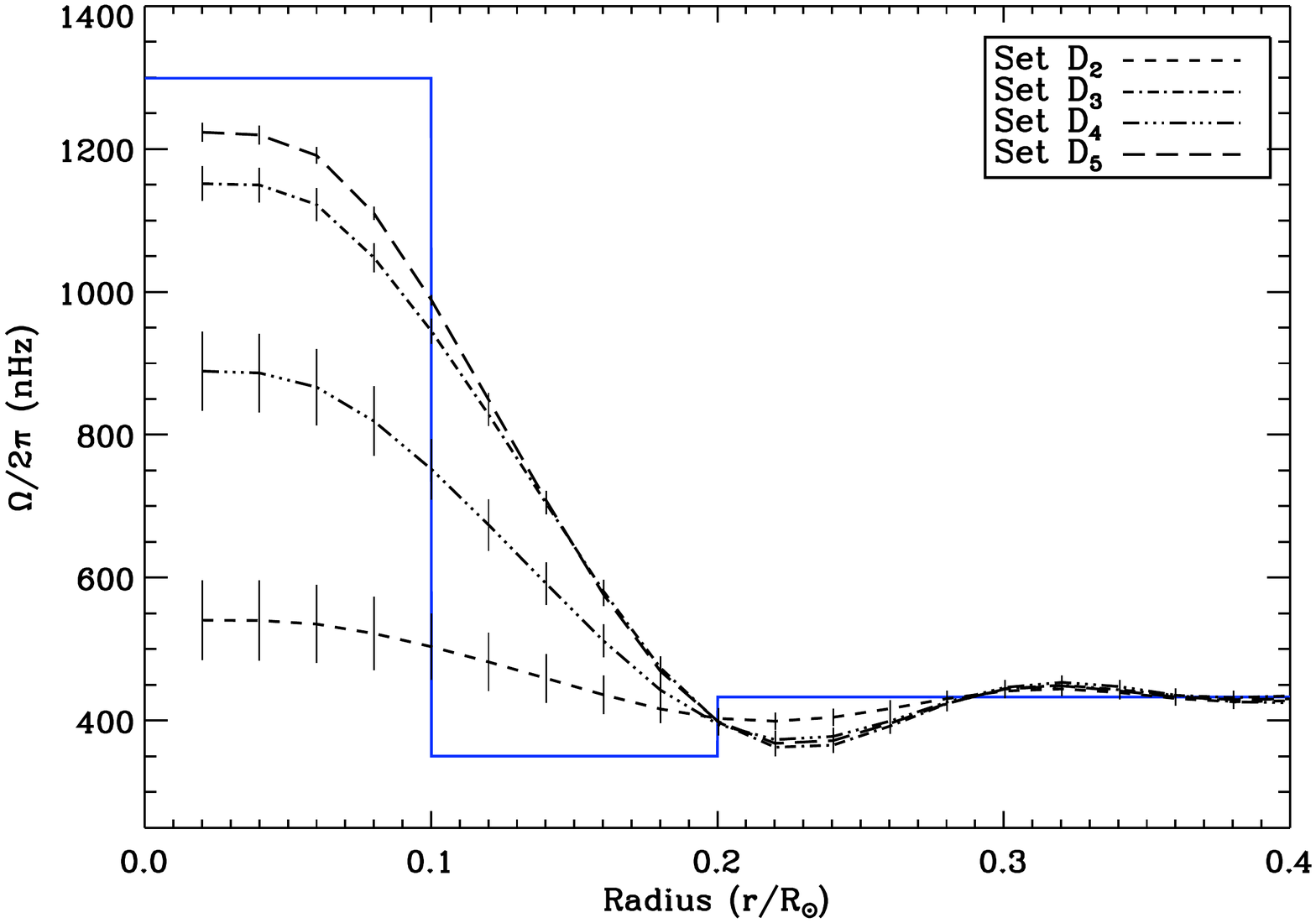}
\caption{Equatorial rotation profiles below 0.4 $R_\odot$ reconstructed from the set $D_2$ (i.e., including the g mode $\ell$=2 $n$=-3 with an error bar of 75 nHz), $D_3$(i.e., including the g mode with an error bar of 7.5 nHz), $D_4$ (i.e., including eight g modes with an error bar of 75 nHz) and $D_5$ (i.e., including eight g modes with an error bar of 7.5 nHz) for the step profile. }
\label{step_4}
\end{center}
\end{figure}

\begin{figure}[htbp]
\begin{center}
\includegraphics[height=7cm]{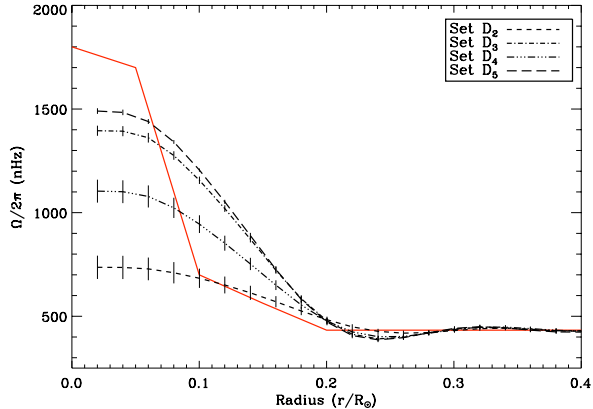}
\caption{Equatorial rotation profiles below 0.4 $R_\odot$ reconstructed from the set $D_2$ (i.e., including the g mode $\ell$=2 $n$=-3 with an error bar of 75 nHz), $D_3$(i.e., including the g mode with an error bar of 7.5 nHz), $D_4$ (i.e., including eight g modes with an error bar of 75 nHz) and $D_5$ (i.e., including eight g modes with an error bar of 7.5 nHz) for the smooth profile. Same legend as Fig.\ref{step_4}.}
\label{smooth_4}
\end{center}
\end{figure}

\begin{figure}[htb*]
\begin{center}
\includegraphics[width=9cm]{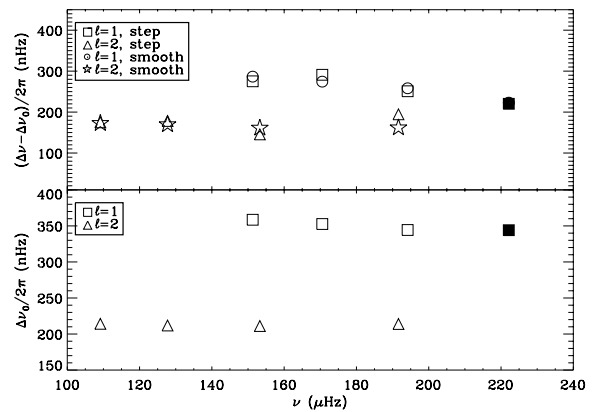}
\caption{Splittings differences of g modes between the rigid profile and on one hand, the step profile  and on the other hand, the smooth profile (top). We have only represented the eight g modes used in the inversions. The filled square corresponds to the g-mode candidate $\ell$=2 n=-3. We have also plotted the splittings of these modes for the rigid profile (bottom).}
\label{dif_split}
\end{center}
\end{figure}

When one g mode ($\ell$=2, n=-3 around 220 $\mu$Hz) is added to the p-mode data set, as in the case of sets {\Dtwo} and {\Dthree}, the inversion results improve below 0.1 $\RSun$ (see Fig.\ref{step_4} and Fig.\ref{smooth_4}), but there is not substantial improvement around  0.2 $\RSun$. The match between the artificial rotational profiles and the profiles estimated from the inversions improves when the error assigned to the g mode is reduced. Unlike the case of the inversion of only p modes, the resolution kernels at 0.08 $\RSun$ significantly improves when adding one g mode (see Fig.\ref{ave_0_08}), in particular when the observational uncertainty falls to 7.5 nHz. The resolution kernel at 0.16 $\RSun$ does not change (or slightly) with the addition of the g mode (with an error bar of 75 nHz), when compared to the same resolution kernel calculated from the inversion of p modes. Unlike the splittings calculated for p modes, the differences in the frequency splittings calculated  from the three artificial rotational profiles for the g modes could be larger than 200 nHz (see Fig.\ref{dif_split}). It is shown in this figure that the differences of splittings calculated for a rigid profile and for the other two simulated rotation profiles, are around 200 nHz. This means that following the usual criteria of $\sim$3 $\sigma$ to have a proper detection, a difference of 200 nHz is visible with the modes having an uncertainty of 75 nHz. Therefore, in this condition, it is possible to discriminate between the rigid profile and the other one with a higher rotation rate in the core.

If eight g modes are added to the p-mode data set, as in sets {\Dfour} and {\Dfive}, the inversion results (see Figs.\ref{step_4} and \ref{smooth_4}) in both the level of uncertainties of the estimates and the matching to the proxy rotation profiles are significantly better than those obtained from the inversion with only one g mode. The differences in the frequency splittings calculated from the three artificial rotational profiles for all the eight g modes are significantly larger than the observational uncertainties and hence, new information could be gained as compared to that given by just one g mode. This is particularly important in the presence of noisy data, since the larger the number of g modes, the better the averaging of the unwanted effects of the noise in the data will be. The addition of several g modes helps to better define the resolution kernels below 0.1 $\RSun$ (see the resolution kernel at 0.08 $\RSun$ Fig.~\ref{ave_0_08}), whereas the resolution kernel at 0.16 $\RSun$ does not significantly change by the addition of g modes. In this sense, the inferences about the rotational rate of the core will be significantly improved below  0.1 $\RSun$, where the energy of the g modes is maximum. Very high frequency p modes (above 2.5 mHz) for $\ell$=1 and 2 should be characterized to better define the region between 0.15 and 0.25 $\RSun$ \citep{2008arXiv0802.1510G}

\subsection{Inversion of real data}

We have used inversions to study the compatibility of present p-mode frequency splittings with the splittings estimated by \citet{STCGar2004} for the $\ell$=2, n=-3 g mode. The p-mode set corresponds to the mode set used in {\Done}, where
the splittings  correspond to those calculated by \citet{kor2005} for 2088 day-long MDI time-series. In \citet{STCGar2004}, three scenarios were proposed to explain the detected pattern around 220 $\mu$Hz, with two possible values for the splittings, namely 300 nHz, if this were a detection of two modes (a combination of an $\ell$=2 and a $\ell$=5), and 600 nHz, if all the visible components correspond to the same mode (the $\ell$=2 $n$=-3 which implies an inclined core rotation axis).

\begin{figure}[h*]
\begin{center}
\includegraphics[height=7cm]{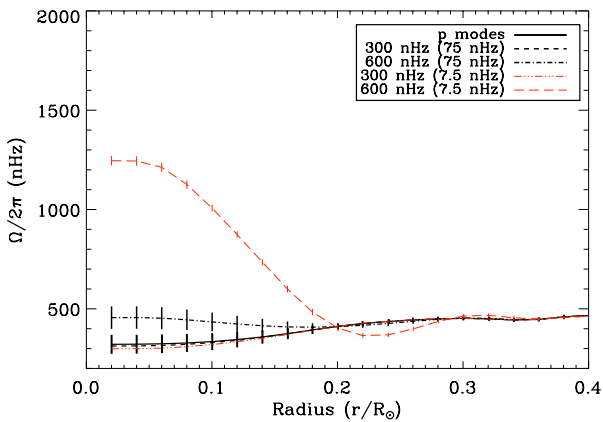}
\caption{Equatorial rotation profiles reconstructed with the real data as explained in the text and adding the g-mode candidate $\ell$=2 $n$=-3 with two different splittings (300 and 600 nHz) combined with two different error bars (75 and 7.5 nHz).}
\label{realg}
\end{center}
\end{figure}

Five inversions are carried out (see Fig.\ref{realg}), namely one inversion including only p modes and other four containing one g mode, but with different estimates of the frequency splitting (300 and 600 nHz) and two observational uncertainties (75 and 7.5 nHz). As it was illustrated in Figs.~\ref{rot_p}, \ref{dif_split1} and \ref{ave_p}, there is no sensitivity of the observed p modes to the dynamics of the inner solar core. Different rotational profiles below 0.2 $\RSun$ are obtained for the different combinations of the value of the introduced g-mode splitting and its corresponding uncertainty. In all cases, 
these values are compatible with the data calculated for the p modes in the sense that the inversions are unchanged and stable (e.g. the inversion does not show any oscillatory behaviour) above 0.2 $\RSun$ when the g mode is added to the data set. In Fig.\ref{step_4}, an oscillation around 0.2 $\RSun$ appears with the artificial data when g modes are associated to an error bar of 7.5 nHz. This is also observed with the real data. This is an artefact of the inversion. The rotation profile obtained using the highest value of the g-mode splitting (600 $\pm$7.5 nHz) proposed by \citet{STCGar2004}, gives a rate in the inner core that is compatible with the result from the dipole analysis carried out by \citet{2007Sci...316.1591G}.
 
To quantify the compatibility between the p-mode data and the g-mode candidate, we have calculated the normalised residuals for all the observed modes ($\ell, m, n$) defined by  $(\delta \nu_{\rm data} - \delta \nu_{\rm inv})/{\sigma}$, where $\delta \nu_{\rm data}$ is the value of the splitting in the data, $\delta \nu_{\rm inv}$, the value corresponding to the rotation profile obtained with the inversion and $\sigma$, the error bar associated to the splitting of the mode. Table~\ref{table2} gives the mean value of these residuals for the low-degree p modes ($\ell \le$3, below 2 mHz) and the g-mode candidate for the four inversions  (the two values for the splittings and the two values for the uncertainty for the g mode). We can see that the difference of splitting between the real data and the results of the forward problem on the inferred rotation profile is less than 1.5 $\sigma$ for the p modes. However, for the g-mode candidate, this difference goes up to $\sim$3 $\sigma$. This is due to the fact that the rotation profile has some uncertainties that have an impact on the splittings calculated. Globally, the results with the g-mode candidate are compatible with the information contained in the observed p-mode data.

\begin{table*}[ht]
  \begin{center} 
\caption{Normalised residuals for p modes ($\ell$ =1, 2 and 3 and $\nu \le$ 2 mHz) and the g-mode candidate.}
\vspace{1em}
\renewcommand{\arraystretch}{1.2}
\begin{tabular}[h]{lcclclclclcl}
\hline
&  600 nHz, $\epsilon$ = 75 nHz & 600, nHz $\epsilon$ = 7.5 nHz & 300 nHz, $\epsilon$ = 75 nHz & 300 nHz, $\epsilon$ = 7.5 nHz\\
\hline
$\ell$ = 1 & -0.13 & - 0.69 & -0.03 & -0.02\\
\hline 
$\ell$ = 2  & 0.34 & -0.42 & 0.47 & 0.49\\
\hline 
$\ell$ = 3 &  -1.13 & -1.23 & -1.10 & -1.10\\
\hline
$\ell$ = 2, $n$ = -3 & 3.36 & 2.16 & -0.05 & -0.04\\
\hline 
      \end{tabular}
    \label{table2}

  \end{center}
\end{table*}

\section{Conclusions}

In this paper, we have studied how the inversion of several artificial rotation profiles could be improved when g modes are added to the present set of observed p modes. The introduction of one g mode --the candidate $\ell$=2, n=-3-- significantly improves
 the solution in the inner core (below 0.1 $\RSun$), when compared to that obtained using only p modes. It gives the general trend of the solar core rotation but there is not accurate information on the profile itself. If more g modes are added to the inversion data set (four $\ell$=1 and four sectoral $\ell$=2), the result in terms of accuracy and error propagation, improves compared with the inversion including only 1 g mode. However there is still information missing in the region between 0.1 - 0.2 $\RSun$, where the energy of the g modes is significantly lower than in the region below 0.1 $\RSun$. The information given by the p modes is negligible due to the lack of sensitivity to these depths, the high level of uncertainties we have in their determination and the noise present in the data.

Finally, for the real data, the rotation profile obtained using the highest value of the g-mode splitting gives a rate in the inner core that is compatible with the result obtained with an independent technique by \citet{2007Sci...316.1591G}, if we put an error bar of 7.5 nHz. Moreover, we obtained a limit down to which we can trust the inversion of the real data. All the values proposed for the splittings of the $\ell$=2, n=-3 g-mode candidate are compatible with the splittings calculated for the p modes. Indeed, having in mind that the small oscillation is related to the inversion and not to the data, the addition of  the g-mode candidate with different values for the splitting and their uncertainty, does not change the estimated profile above 0.2 $\RSun$. 

\begin{acknowledgements}
This work has been partially funded by the grant AYA2004-04462 of the Spanish Ministry of Education and Culture and partially supported by the European Helio- and Asteroseismology Network
(HELAS\footnote{http://www.helas-eu.org/}), a major international collaboration funded by the European Commission's Sixth Framework Programme.

\end{acknowledgements}

\bibliographystyle{aa} 
\bibliography{/Users/smathur/Documents/BIBLIO_sav}  

\end{document}